\documentclass[journal]{IEEEtran}

\usepackage{cite}
\usepackage{url}
\usepackage{graphicx}
\usepackage{color}

\usepackage{amsmath}
\usepackage{amsthm,amsfonts}
\usepackage{pgfplots}

\usepackage{tikz}
\usepackage[customcolors]{hf-tikz}
\usepgfplotslibrary{colorbrewer}
\usetikzlibrary{%
  3d,
  arrows,%
  shapes.misc,
  shapes.arrows,%
  chains,%
  matrix,%
	fit,%
  positioning,
  scopes,%
  decorations.pathmorphing,
  shadows,%
  decorations.markings,
  shapes.geometric,
  arrows.meta,bending,
	patterns,
	intersections, 
	pgfplots.fillbetween
}

\usepackage{tikz-3dplot}
\usepackage[pdfusetitle]{hyperref}
\hypersetup{
    colorlinks,
    linkcolor={red!50!black}, 
    citecolor={blue!50!black},
    urlcolor={blue!50!black}
}

\usepackage{cellspace}
\providecommand*{\M}[1]{\mathbf#1}
\providecommand*{\tM}[1]{\tilde{\mathbf#1}}
\providecommand*{\wtM}[1]{\widetilde{\mathbf#1}}
\providecommand*{\bM}[1]{\bar{\mathbf#1}}
\providecommand*{\wbM}[1]{\bar{\mathbf#1}}
\providecommand*{\mrm}[1]{\mathrm{#1}}
\providecommand*{\unit}[1]{\ensuremath{\mrm{\,#1}}}

\providecommand*{\V}[1]{\boldsymbol#1}
\providecommand*{\UV}[1]{\hat{\boldsymbol#1}}
\providecommand*{\T}[1]{\mathrm{#1}}
\DeclareMathAccent{\ring}{\mathalpha}{operators}{"17}

\providecommand*{\unit}[1]{\ensuremath{\mrm{\,#1}}}
\providecommand*{\eu}{\ensuremath{\mrm{e}}}

\providecommand*{\ju}{\ensuremath{\mrm{j}}}
\renewcommand{\Re}{\operatorname{Re}}	
\renewcommand{\Im}{\operatorname{Im}}	

\providecommand*{\diffV}{\operatorname{dV}\!}

\newcommand{\ie}{\textit{i.e.}\/, }
\newcommand{\eg}{\textit{e.g.}\/, }
\newcommand{\cf}{\textit{cf.}\/, }

\newcommand{\trans}{\text{T}}
\newcommand{\herm}{\text{H}}

\newcommand{\maximize}{\mrm{maximize}}

\newcommand{\subto}{\mrm{subject\ to}}

\newcommand{\Rs}{R_\mrm{s}}
\newcommand{\basv}{\V{\psi}}
\newcommand{\reg}{\varOmega}

\newcommand{\ellx}{\ell_\T{x}}
\newcommand{\elly}{\ell_\T{y}}

\newcommand{\Ri}{\M{G}}

\colorlet{dpurple}{blue!50!red}
\colorlet{dblue}{blue!50!black}
\colorlet{dgreen}{green!50!black}
\colorlet{dred}{red!50!black}
\colorlet{dyellow}{yellow!50!black}
\colorlet{dorange}{orange!50!black}
\definecolor{metal}{RGB}{218,165,32}
\definecolor{diel}{RGB}{1,165,32}
\definecolor{antenna}{RGB}{100,150,162}
\definecolor{breg}{rgb}{0.2,0.6,0.8}%
\definecolor{preg}{rgb}{0.8,0.2,0.2}%
\definecolor{reg}{RGB}{218,165,32}

\graphicspath{{figures/}{tikz/}}

\makeatother

\title{Physical Limits and Optimal Synthesis\\ of Beyond Diagonal Anomalous Scatterers}

\author{
Mats Gustafsson
\thanks{This work was supported by the Swedish Research Council SEE-6GIA and SSF Sabbatical.}
\thanks{M. Gustafsson is with Electrical and Information Technology, Lund University, Lund, Sweden, (e-mail: mats.gustafsson@eit.lth.se).}
}

\tdplotsetmaincoords{70}{140}
\pgfmathsetmacro{\wx}{0.8} 
\pgfmathsetmacro{\wy}{3.1} 
\pgfmathsetmacro{\wz}{0.4} 
\pgfmathsetmacro{\fx}{0} 
\pgfmathsetmacro{\fy}{1}
\pgfmathsetmacro{\fz}{0.4}
\pgfmathsetmacro{\fw}{0.2}
\pgfmathsetmacro{\hx}{2} 
\pgfmathsetmacro{\hy}{2.7}
\pgfmathsetmacro{\hz}{1}
\pgfmathsetmacro{\d}{0.2}
\tikzset{%
		grid/.style={very thin,gray},
		axis/.style={->,white,thin},
		cube/.style={fill=metal,fill opacity=0.5,draw=blue!50!black},
    horn/.style={fill=black!50!blue,fill opacity=0.5,draw=blue!50!black},
		cube hidden/.style={fill=black!50!blue,fill opacity=0.5,draw=blue!50!black},
    xyplane/.style={canvas is xy plane at z=#1,very thin}    
    }
\tikzset{>=latex}

\begin{document}

\maketitle
\begin{abstract}
Realizing metasurfaces for anomalous scattering is fundamental to designing reflector arrays, reconfigurable intelligent surfaces, and metasurface antennas. However, the basic cost of steering scattering into non-specular directions is not fully understood.
This paper derives tight physical bounds on anomalous scattering using antenna array systems equipped with non-local matching networks. The matching networks are explicitly synthesized based on the solutions of the optimization problems that define these bounds. Furthermore, we analyze fundamental limits for metasurface antennas implemented with metallic and dielectric materials exhibiting minimal loss within a finite design region.
The results reveal a typical 6dB reduction in bistatic radar cross section (RCS) in anomalous directions compared to the forward direction. Numerical examples complement the theory and illustrate the inherent cost of achieving anomalous scattering relative to forward or specular scattering for canonical configurations.
\end{abstract}

%
\section{Introduction}

Metasurfaces are widely used to engineer electromagnetic scatterers and antennas by patterning surfaces with spatially varying subwavelength elements, enabling control of the scattered fields in anomalous (non-specular) directions and with different beam patterns~\cite{Glybovski+etal2016,Minatti+etal2017,Kosulnikov+etal2024,Vuyyuru+etal2023,Maci2024}.  Specific examples include: reflector arrays, which are planar structures designed to, \eg mimic the functionality of curved reflectors~\cite{Nayeri+etal2018}. Metasurface antennas can generate complex radiation patterns from a single flat aperture excited by a single feed~\cite{Minatti+etal2014,Teodorani+etal2024}. More recently, Reconfigurable Intelligent Surfaces (RIS) have emerged as a paradigm, introducing dynamic control over the local response of the metasurface elements~\cite{Vuyyuru+etal2025}. By tuning the surface in real time, RIS can steer reflected or scattered fields in desired directions, adaptively optimizing wireless communication links~\cite{Li+etal2024,Nerini+etal2024}. 

This paper focuses on the fundamental limits of passive metasurface scatterers made of metallic and dielectric materials. Specifically, we investigate the extent to which passive surfaces, \ie constructed by passive materials, can shape the scattered field under given constraints. Our goal is to characterize the ultimate performance boundaries of metasurface-based reflectors in terms of beam shaping.
Explicit synthesis techniques have demonstrated that full anomalous reflection or refraction is difficult using dielectric and passive metasurfaces~\cite{Asadchy+etal2017,Mohammadi+Alu2016}. In this work, we formulate the scattering problem as an optimization problem over all physically admissible passive realizations. To make this tractable, we relax the problem to a quadratically constrained quadratic program (QCQP)~\cite{Park+Boyd2017} over the space of induced currents, enabling the derivation of fundamental performance limits~\cite{Gustafsson+etal2020,Gustafsson+Nordebo2013,Chao+etal2022,Gustafsson+Capek2019,Gustafsson+etal2015b,Jelinek+Capek2017,Kuang+etal2020,Abdelrahman+Monticone2023}.

We present formulations for both given antennas and arbitrary design regions, allowing for a comprehensive analysis of performance limits under different design constraints. For antennas, the bounds are determined for all passive matching networks and for regions over all possible passive antenna realizations based on metals and dielectrics with some losses. Furthermore, the computed optimal currents~\cite{Gustafsson2024b} are employed to synthesize a non-local (i.e., beyond-diagonal) matching network for the fixed antenna structures, and a non-local material distribution for the arbitrary regions. These synthesized solutions offer insight into the physical mechanisms required to achieve optimal scattering under passivity constraints.

The presented approach builds on~\cite{Gustafsson+etal2020,Abdelrahman+Monticone2023} and determines the maximum achievable scattering in a specified direction or, more generally, the optimal field pattern within a designated near-field or far-field observation region, using a passive antenna structure. The results are illustrated for rectangular design regions in free space or placed above an infinite ground plane. In these examples, a reduction in anomalous scattering compared to forward scattering, by approximately a factor of $4$ (corresponding to $-6\unit{dB}$), is observed. This reduction is quantitatively explained by the explicit solution, which yields a simple closed-form estimate for these configurations, offering intuitive insight into the fundamental limitations imposed by passivity, geometry, and materials.

The array antenna setup is presented in Sec.~\ref{S:AntennaModel}. Maximal scattering from an array is derived in Sec.~\ref{S:MaxScattAntenna}, followed by synthesis of a non-local matching network for optimal scattering in Sec.~\ref{S:Synthesis}. Fundamental limits on all scatterers realized within a design region are presented in Sec.~\ref{S:MaxScatt} with simplifications for plane waves and far fields in Sec.~\ref{S:PlanewaveScatt}. 
Sec.~\ref{S:NumRes} presents numerical results. The paper is concluded in Sec.~\ref{S:Conclusions}.

\section{Antenna and Scatterer Model}\label{S:AntennaModel}
\begin{figure}
    \centering
    \includegraphics[trim={2cm 0 0 0},clip,width=0.95\linewidth]{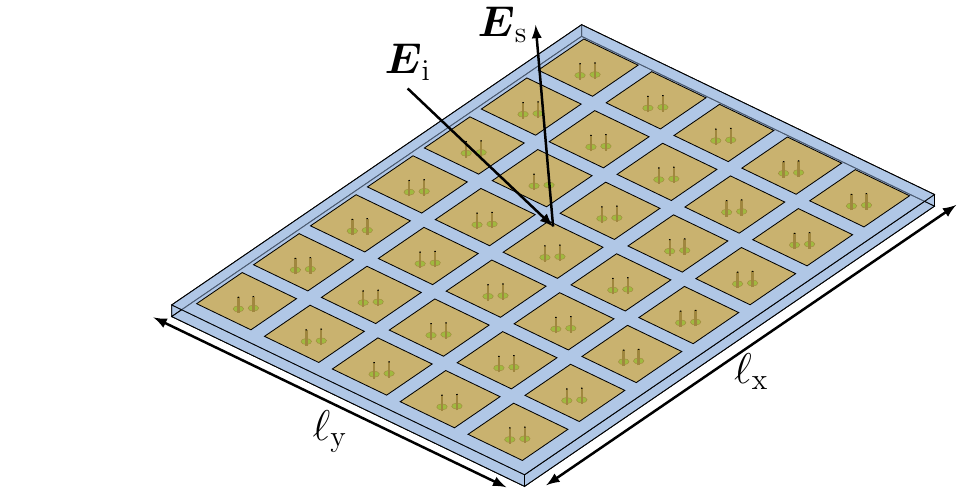}   
    \caption{Illustration of an antenna scatterer consisting of $5\times 7$ dual polarized patch elements.}
    \label{fig:antennascatt}
\end{figure}

Consider scattering of an antenna array structure as depicted in Fig.~\ref{fig:antennascatt}. The antenna is of arbitrary shape contained in a region $\reg$ and assumed to have $N_\T{p}$ ports. Scattering of the antenna depends on the termination of the antenna ports. Here, we assume a general linear, time-translational invariant, and passive matching network connecting the antenna ports~\cite{Pozar2005}. This matching network is modeled as an impedance matrix $\M{Z}_\T{L}$ with real (resistance) $\Re\{\M{Z}_\T{L}\}$ and imaginary (reactance) $\Im\{\M{Z}_\T{L}\}=\M{X}_\T{L}$ parts~\cite{Pozar2005}. The reactance is assumed arbitrary and the resistance is positive definite (PD) $\Re\{\M{Z}_\T{L}\}\succeq \M{R}_\T{L}\succ\M{0}$. The network is termed reciprocal if $\M{Z}_\T{L}^{\trans}=\M{Z}_\T{L}$ and local (diagonal) of $\M{Z}_\T{L}$ is a diagonal matrix, meaning that the ports are not interconnected through the feed network. Instead, mutual coupling occurs solely via the electromagnetic fields outside the matching network. In contrast, a 'non-local' (or 'beyond-diagonal') load matrix contains off-diagonal terms~\cite{Nerini+etal2024,Li+etal2024}.

An antenna operating as a passive scatterer can be accurately modeled using an Electric Field Integral Equation (EFIE) Method of Moments (MoM) impedance matrix $\M{Z}$ as~\cite{Harrington1968}  
\begin{equation}
   \M{Z} \M{I} = \M{V},
\end{equation}  
where $\M{V}$ is the excitation and $\M{I}$ the induced current on the antenna structure. Here, the current density $\V{J}(\V{r})$ is expanded in basis functions~\cite{Harrington1968} $\basv_n(\V{r})$, \ie
\begin{equation}
    \V{J}(\V{r}) = \sum_n I_n\basv_n(\V{r}),
\end{equation}
where the expansion coefficients $I_n$ are collected in a column matrix $\M{I}$. 
The MoM impedance matrix contains the geometrical and material properties of the antenna as well as the modeling of the antenna ports~\cite{Harrington1968}, but not the matching network. We assume that the excitation is external, such as an incident plane wave from a given direction or the field from a near-field source. 

We decompose the current $\M{I}=[\M{I}_\T{c}\ \M{I}_\T{b}]^{\trans}$ into controllable currents $\M{I}_\T{c}$ belonging to the $N_\T{p}$ feed ports and currents on the background antenna structure $\M{I}_\T{b}$. This block partitions the impedance matrix~\cite{Parhami+etal1977}
\begin{equation}
    \M{Z} = \begin{pmatrix}
        \M{Z}_\T{cc} & \M{Z}_\T{cb}\\
        \M{Z}_\T{bc} & \M{Z}_\T{bb}
    \end{pmatrix}
\end{equation}
and similar for the excitation $\M{V}=[\M{V}_\T{c}\ \M{V}_\T{b}]^{\trans}$. Elimination of the background current $\M{I}_\T{b}=\M{Z}_\T{bb}^{-1}\M{V}_\T{b}-\M{Z}_\T{bb}^{-1}\M{Z}_\T{bc}\M{I}_\T{c}$ construct an antenna impedance matrix 
\begin{equation}
    \wtM{Z}\M{I}_\T{c}=(\M{Z}_\T{cc}-\M{Z}_\T{cb}\M{Z}_\T{bb}^{-1}\M{Z}_\T{bc})\M{I}_\T{c}
    =\M{V}_\T{c}-\M{Z}_\T{cb}\M{Z}_\T{bb}^{-1}\M{V}_\T{b}
    =\wtM{V},
    \label{eq:MoMsub}
\end{equation}
where we use the symbol tilde to denote the antenna impedance matrix. The matching network is restricted to the $N_\T{p}$ ports belonging to the controllable region. Matching networks can be modeled in many ways, including scattering, impedance, and admittance formulations~\cite{Pozar2005}. Here, we investigate what can theoretically be achieved with an optimal matching network and use an impedance matrix to model the class of linear time-invariant matching networks. Let $\M{Z}_\T{L}$ denote the impedance matrix of the matching network added to $\wtM{Z}$. 

The scattered electric (or magnetic) field at a point $\V{r}$ or in a direction $\UV{r}$ with polarization $\UV{e}$ is modeled as $\M{F}^{\herm} \M{I}$, where $\M{F}$ is a matrix that maps the current to the field at the desired location, see App.~\ref{S:MoM}. Note that the same notation is used here for both near and far fields. The matrix $\M{F}$ is normalized such that the expression $|\M{F}^{\herm} \M{I}|^2$ is proportional to radiated or scattered power of interest. 
Decomposing the matrix $\M{F}$ into the controllable and background regions
\begin{equation}
    \M{F}^{\herm}\M{I} 
    = \M{F}_\T{c}^{\herm}\M{I}_\T{c} + \M{F}_\T{b}^{\herm}\M{I}_\T{b}
    = (\M{F}_\T{c}-\M{Z}_\T{bc}^{\herm}\M{Z}^{-\herm}_\T{bb}\M{F}_\T{b})^{\herm}\M{I}_\T{c}
    +\M{F}_\T{b}^{\herm}\M{Z}^{-1}_\T{bb}\M{V}_\T{b}
    \label{eq:Fsub}
\end{equation}
is used to model the radiation from the antenna ports and the background separately. 

\begin{figure}
    \centering
     \includegraphics[trim={2cm 0 0 0},clip,width=0.95\linewidth]{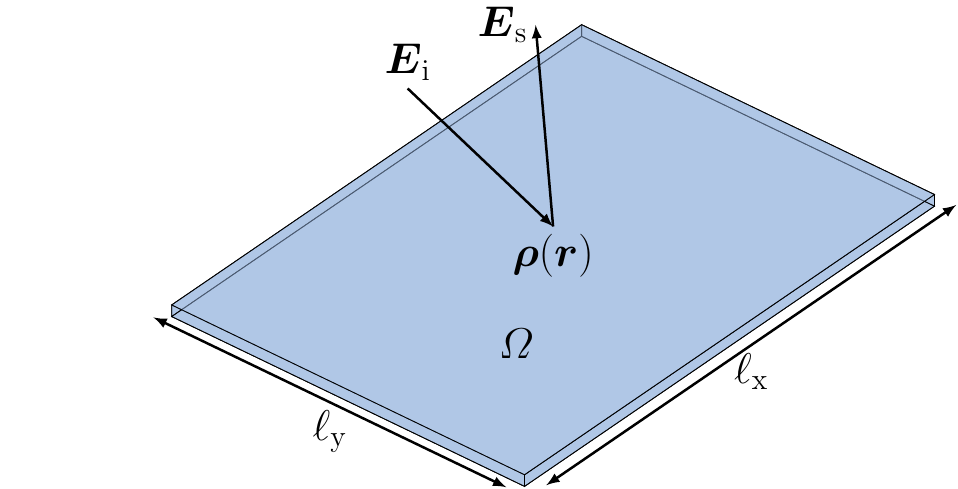}   
    \caption{The antenna in Fig.~\ref{fig:antennascatt} is contained with in a design rectangular region $\reg$ modeled by the complex resistivity $\V{\rho}(\V{r})$ with side lengths $\ellx$ and $\elly$.}
    \label{fig:antennascattreg}
\end{figure}

The antenna model~\eqref{eq:MoMsub} assumes a fixed antenna structure together with a matching network, see Fig.~\ref{fig:antennascatt}. To instead model all antenna structures that can be realized in a design region $\reg$ using a set of materials such as copper or plastic, we assume that the scatterer is realized by materials with non-negligible loss, such as metals and dielectrics, or vacuum, see Fig.~\ref{fig:antennascattreg}. The material is characterized by an anisotropic complex resistivity distribution $\V{\rho}(\V{r})$, where the real part (associated with material resistance) satisfies the constraint $\Re\{\V{\rho}(\V{r})\} \succeq \V{\rho}_\T{r} \succ \V{0}$,
for some minimum resistance level $\V{\rho}_\T{r}$, while the imaginary part (reactive component) is arbitrary. Importantly, the determined bounds remain valid even when allowing for non-local reactivity, thereby encompassing a wide range of possible metamaterial and metasurface designs. The material modeling in this paper is restricted to dielectric materials, \ie arbitrary linear constitutive relations connecting the electric and electric displacement fields~\cite{VanBladel2007}. However, homogenized materials models of magnetic material based on meta-elements constructed by metal and dielectrics are included in the model. 

Note that the material model is not diagonal when using global or divergence conforming basis functions~\cite{Jin2010}. Nevertheless, in this work, we continue to refer to such cases as 'local' or 'diagonal' for consistency, even when the material matrix is not strictly diagonal.

\section{Maximal scattering for a given antenna}\label{S:MaxScattAntenna}
The port current $\tM{I}$ due to an excitation $\wtM{V}$ is obtained from the system equation $(\wtM{Z}+\M{Z}_\T{L}) \tM{I} = \wtM{V}$ with $\wtM{Z}$ from~\eqref{eq:MoMsub}, $\M{Z}_\T{L}$ being the matching network, and $\tM{I}$ denoting the current in the controllable region (feed) for the antenna system including the matching network. The maximum power that can be delivered, over all possible matching networks, for a given excitation, is determined by maximizing
\begin{equation}
    \left|\M{F}^{\herm}\M{I}\right|^2
    =\wtM{I}^{\herm}\wtM{F}\wtM{F}^{\herm}\wtM{I}
    +\Re\{\wtM{I}^{\herm}\tM{f}\}+\tilde{f}
\end{equation}
for all networks $\M{Z}_\T{L}$, \ie antenna systems satisfying $(\M{Z}_\T{L}+\wtM{Z})\wtM{I}=\wtM{V}$, and with the parameters $\tM{f},\tilde{f}$ expressed using~\eqref{eq:Fsub}. This is conveniently written as an optimization problem
\begin{equation}
\begin{aligned}
& \maximize_{\M{Z}_\T{L}} &&  \tM{I}^{\herm}\M{F}\M{F}^{\herm}\tM{I}+\Re\{\tM{I}^{\herm}\tM{f}\}+\tilde{f} \\
& \subto && (\wtM{Z}+\M{Z}_\T{L})\tM{I}
=\wtM{V}, 
\end{aligned}
\label{eq:MaxU1}
\end{equation}
where the optimization is over all admissible loads $\M{Z}_\T{L}$, \ie $\Re\{\M{Z}_\T{L}\}\succeq \M{R}_\T{L}$. Relax the optimization problem by multiplying the constraint with the currents $\tM{I}^{\herm}$ 
\begin{equation}
\begin{aligned}
& \maximize_{\M{Z}_\T{L},\tM{I}} &&  \tM{I}^{\herm}\wtM{F}\wtM{F}^{\herm}\tM{I}+\Re\{\tM{I}^{\herm}\tM{f}\}+\tilde{f} \\
& \subto && \tM{I}^{\herm}(\M{Z}_\T{L}+\wtM{Z})\tM{I}=\tM{I}^{\herm}\wtM{V}.
\end{aligned}
\label{eq:MaxU2}
\end{equation}
Further using $\Re\{\M{Z}_\T{L}\}\succeq\M{R}_\T{L}$ and only keeping the real part of the constraint produces the QCQP
\begin{equation}
\begin{aligned}
& \maximize_{\tM{I}} &&  \tM{I}^{\herm}\wtM{F}\wtM{F}^{\herm}\tM{I}+\Re\{\tM{I}^{\herm}\tM{f}\}+\tilde{f} \\
& \subto && \tM{I}^{\herm}(\M{R}_\T{L}+\wtM{R})\tM{I}\leq \Re\{\tM{I}^{\herm}\wtM{V}\}, 
\end{aligned}
\label{eq:MaxU3}
\end{equation}
where $\wtM{R}=\Re\{\wtM{Z}\}$ and the optimization is solely over the currents $\tM{I}$.

Solution of this relaxed problem~\eqref{eq:MaxU3} always produces a value greater or equal to the value of the original problem~\eqref{eq:MaxU1}, \ie any viable current in~\eqref{eq:MaxU1} is a potential current in~\eqref{eq:MaxU3}. This QCQP~\eqref{eq:MaxU3} is easily solved using duality~\cite{Gustafsson+etal2020}
\begin{equation}
    U = \min_{\nu\geq \nu_1}
    \frac{1}{8}(\tM{f}+\nu\wtM{V})^{\herm}\left(
    \nu\M{R}_\T{A}-\wtM{F}\wtM{F}^\herm
    \right)^{-1}(\wtM{f}+\nu\wtM{V})+\tilde{f}
    \label{eq:dualA}
\end{equation}
with $\nu_1=\wtM{F}^{\herm}\M{R}_\T{A}^{-1}\wtM{F}$ and $\M{R}_\T{A}=\wtM{R}+\M{R}_\T{L}$ is used to simplify the notation. The dual problem is convex and \eg minimized using a line search algorithm~\cite{Boyd+Vandenberghe2004}. Using Sherman-Morrison formula~\cite{Golub+Loan2013} for analytical inversion is further used to reduce the computational complexity of the dual problem~\eqref{eq:dualA} 
\begin{equation}
    U = \min_{\nu\geq \nu_1}
    \frac{(\tM{f}+\nu\wtM{V})^{\herm}}{8\nu}\left(
    \M{R}_\T{A}^{-1}+\frac{\M{R}_\T{A}^{-1}\wtM{F}\wtM{F}^\herm\M{R}_\T{A}^{-1}}{\nu-\wtM{F}^\herm\M{R}_\T{A}^{-1}\wtM{F}}
    \right)(\tM{f}+\nu\wtM{V})+\tilde{f},
    \label{eq:dualB}
\end{equation}
where all matrix multiplication can be pre-calculated. 

Solution of the bound~\eqref{eq:dualB} determines the dual parameter $\nu=\nu_\T{o}$, upper bound on $U$, and the optimal current $\M{I}_\T{o}$ 
\begin{equation}
    \tM{I}_\T{o}
    =\frac{1}{2}(\nu_\T{o}\M{R}_\T{A}-\wtM{F}\wtM{F}^{\herm})^{-1}
    (\tM{f}+\nu_\T{o}\wtM{V}).
    \label{eq:Iopt}
\end{equation}
The QCQP~\eqref{eq:MaxU3} with a PD objective functional and a PD constraint has no dual gap~\cite{Beck+Eldar2006} and the optimal current $\tM{I}_\T{o}$ satisfies the constraint~\eqref{eq:MaxU3} in equality, \ie 
\begin{equation}
    \tM{I}_\T{o}^{\herm}\M{R}_\T{A}\tM{I}_\T{o}
    =\tM{I}_\T{o}^{\herm}(\M{R}_\T{L}+\wtM{R})\tM{I}_\T{o}
    =\Re\{\tM{I}_\T{o}^{\herm}\wtM{V}\}.
    \label{eq:IoptEq}
\end{equation}
This equality is next used to synthesize a matching network $\M{Z}_\T{L}$ satisfying~\eqref{eq:MaxU1}. 

\section{Network synthesis }\label{S:Synthesis}
The bound from the dual problem~\eqref{eq:dualA} solves the relaxed formulation~\eqref{eq:MaxU3}; however, it is not immediately clear whether it also solves the original problem~\eqref{eq:MaxU1}. To address this, we need to explicitly synthesize a matching network $\M{Z}_\T{L}$. Notably, the reactance $\M{X}_\T{L}$ can be treated as a free parameter during the synthesis of the matching network, while the resistance is set to its minimal achievable value $\M{R}_\T{L}$.

To synthesize a reactance matrix $\M{X}_\T{L}=\M{X}_\T{A}-\wtM{X}$ solving~\eqref{eq:MaxU1}, we return to the MoM system with the added matching impedance~\eqref{eq:MoMsub} 
\begin{equation}
    (\M{R}_\T{A}+\ju\M{X}_\T{A})\tM{I}_\T{o}
    =(\wtM{R}+\M{R}_\T{L}+\ju(\wtM{X}+\M{X}_\T{L})\tM{I}_\T{o}=\wtM{V}
    \label{eq:X1}
\end{equation}
and the given optimal current $\tM{I}_\T{o}$~\eqref{eq:Iopt} and excitation $\wtM{V}$. 
Here, all quantities are known except the reactance $\M{X}_\T{L}$ or equivalently $\M{X}_\T{A}$. It is sufficient to find one solution. Searching for rank one solutions $\M{X}_\T{A}=\pm\M{Y}\M{Y}^{\herm}$ simplifies the problem~\eqref{eq:X1} to
\begin{equation}
    \pm\ju\M{Y}\M{Y}^{\herm}\tM{I}_\T{o}
    =\wtM{V}-\M{R}_\T{A}\tM{I}_\T{o}
    \label{eq:X3}    
\end{equation}
which by multiplying with $\tM{I}_\T{o}^{\herm}$ reduces to
\begin{equation}
    \ju\tM{I}_\T{o}^{\herm}\M{Y}\M{Y}^{\herm}\tM{I}_\T{o}
    =\pm\tM{I}_\T{o}^{\herm}\wtM{V}\mp\tM{I}_\T{o}^{\herm}\M{R}_\T{L}\tM{I}_\T{o}
    =\pm\ju\delta
\end{equation}
for some real-valued parameter $\delta=\Im\{\tM{I}_\T{o}^{\herm}\wtM{V}-\tM{I}_\T{o}^{\herm}\M{R}_\T{A}\tM{I}_\T{o}\}$, where~\eqref{eq:IoptEq} is used to show that $\delta$ is real-valued.
Choose the sign $\pm$ such that $\pm\delta> 0$ giving $ \M{Y}^{\herm}\tM{I}_\T{o} = \sqrt{|\delta|}$
and by reinserting into~\eqref{eq:X3}
\begin{equation}
    \M{Y} = \mp\ju\frac{\wtM{V}-\M{X}_\T{A}\tM{I}_\T{o}}{\sqrt{|\delta|}},
\end{equation}
where $\pm\ju$ can be dropped for the construction of $\M{X}_\T{A}$.

The reactance is finally determined by subtraction of the antenna reactance matrix from $\M{X}_\T{A}=\pm\M{Y}\M{Y}^{\herm}$, \ie
\begin{equation}
    \M{X}_\T{L}
    =\M{X}_\T{A}-\wtM{X}
    =\pm\M{Y}\M{Y}^\herm-\wtM{X},
    \label{eq:OptZL}
\end{equation}
which in the general case is identified as a non-reciprocal and non-local (beyond diagonal) network. For the special case $\delta=0$, it is sufficient to use $\M{X}_\T{L}=-\wtM{X}$.

The network synthesis~\eqref{eq:OptZL} demonstrates that the bound obtained from the dual problem~\eqref{eq:dualA} is tight, in the sense that there exists a matching network $\M{Z}_\T{L}$ that solves the original problem~\eqref{eq:MaxU1}. This synthesized network may, however, be complex, challenging to realize, and potentially offer, \eg inferior bandwidth properties. Alternative realizations that are easier to implement might also exist.
The practical realization of these matching networks is not the focus of this paper; instead, the objective is to establish fundamental bounds on what can theoretically be achieved using idealized circuits.

The analysis so far has considered given antenna structures with a finite number of antenna ports. In the next step, we extend the framework to encompass all possible passive antennas that can be realized using dielectrics and metals within a specified design region.

\section{Maximal scattering for any antenna}\label{S:MaxScatt}
The bound~\eqref{eq:dualA} and network synthesis~\eqref{eq:OptZL} are derived for a given antenna geometry due to the limited control of the currents outside the antenna ports, see Fig.~\ref{fig:antennascatt}. An upper bound on the scattering of all antennas designed within a design region $\reg$, see Fig.~\ref{fig:antennascattreg}, is found by letting all currents be controllable and considering some minimal material losses. This is analogous to antenna current optimization used to determine fundamental limits~\cite{Gustafsson+Nordebo2013,Gustafsson+Capek2019,Ehrenborg+Gustafsson2018,Jelinek+Capek2017}. This simplifies the optimization problem~\eqref{eq:MaxU1} to the directional scattering problem analyzed in~\cite{Gustafsson+etal2020} and~\cite{Abdelrahman+Monticone2023}. Setting $\tM{f}=\M{0}$, $\tilde{f}=0$ in~\eqref{eq:MaxU1} and using the free-space MoM matrix $\M{Z}_0$ for the region $\reg$ reduces the dual problem~\eqref{eq:dualA} to
\begin{equation}
    U = \min_{\nu\geq \nu_1}
    \frac{\nu^2}{8}\M{V}^{\herm}\left(
    -\M{F}\M{F}^\herm+\nu\M{R}
    \right)^{-1}\M{V}
    \label{eq:dualMoM1}
\end{equation}
with $\nu_1=\M{F}^{\herm}\Ri\M{F}$ as before and $\Ri=\M{R}^{-1}$~\cite{Gustafsson+etal2020} is used for notational simplicity. Using the low-rank structure of $\M{F}$ produces the directional scattering bound~\cite{Gustafsson+etal2020}
\begin{equation}
    U = \frac{1}{8}\left(
    |\M{F}^{\herm}\Ri\M{V}|+\sqrt{\M{V}^{\herm}\Ri\M{V}\M{F}^{\herm}\Ri\M{F}}
    \right)^2
    \label{eq:maxUMoM}
\end{equation}
with the optimal dual parameter $\nu$ given in Sec.~\ref{S:QCQP}.
The matching network for this system is determined as in Sec.~\ref{S:Synthesis} but now simplified to $\M{X}_\T{L}=-\M{X}_0$ and interpreted as a non-local material model. The optimal current~\eqref{eq:Iopt} can be written as a linear combination of two currents
\begin{equation}
    \M{I}_\T{o} 
    =\frac{1}{2}\M{I}_\T{V} + \frac{\alpha}{2} \M{I}_\T{F}    
    =\frac{1}{2}\Ri\M{V} + \frac{\alpha}{2}\Ri\M{F},
    \label{eq:IoptF}
\end{equation}
with weight $\alpha$ given in App.~\ref{S:QCQP}. The currents $\M{I}_\T{V,F}$ can be interpreted as induced currents from excitations $\M{V}$ or $\M{F}$. 

In~\eqref{eq:maxUMoM} the terms containing $\M{V}$ and $\M{F}$ appear in a symmetric form but it is important to note that $\M{V}$ and $\M{F}$ have different physical dimensions, see App.~\ref{S:MoM}. 
To simplify notation and harmonize the expressions, we normalize the excitation $\M{V}$ and radiation $\M{F}$ matrices as
\begin{equation}
    \wbM{V} = \frac{\sqrt{\eta_0}}{|E_0|}\M{V}
    \quad\text{and }
    \wbM{F}=\frac{4\pi}{k}\M{F}
    =2\lambda\M{F}
\end{equation}
giving them the same units, with $E_0$ denoting a normalization of the amplitude of the excitation field, $\lambda=2\pi/k$ the wavenumber, and $k$ the wavenumber. Further, normalizing the scattered power density with the incident power yields
\begin{equation}
    \frac{2U\eta_0}{|E_0|^2} 
    = \frac{1}{16\lambda^2} \left(|\wbM{F}^{\herm}\Ri\wbM{V}|+\sqrt{\wbM{V}^{\herm}\Ri\wbM{V}\wbM{F}^{\herm}\Ri\wbM{F}}
    \right)^2.
    \label{eq:maxUnormMoM}
\end{equation}
All quantities in~\eqref{eq:maxUnormMoM} are positive, and using the Cauchy–Schwarz inequality gives the restriction
\begin{equation}
    \frac{\wbM{V}^{\herm}\Ri\wbM{V}
    \wbM{F}^{\herm}\Ri\wbM{F}}{16\lambda^2}
    \leq
    \frac{2U\eta_0}{|E_0|^2} 
    \leq \frac{\wbM{V}^{\herm}\Ri\wbM{V}
    \wbM{F}^{\herm}\Ri\wbM{F}}{4\lambda^2},
    \label{eq:CSg}
\end{equation}
where it is noted that it is only a factor of 4 between the upper and lower values and that the lower value is obtained in the uncorrelated case $\wbM{F}^{\herm}\Ri\wbM{V}=0$. 

The two symmetric terms in~\eqref{eq:maxUnormMoM} and~\eqref{eq:CSg} have a simple interpretation from the normalized maximal extincted power of an object in $\reg$ illuminated by a field represented by $\wbM{V}$ or $\wbM{F}$~\cite{Gustafsson+etal2020}, \ie
\begin{equation}
\begin{aligned}
& \maximize_{\M{I}} &&  \Re\{\M{I}^{\herm}\wbM{V}\} \\
& \subto && \M{I}^{\herm}\M{R}\M{I}
=\M{I}^{\herm}\wbM{V}, 
\end{aligned}
\label{eq:MaxPt}
\end{equation}
solved by the normalized current $\bM{I}_{\T{V}}=\Ri\wbM{V}$, similar to the currents in~\eqref{eq:Iopt}. The mixed term $\bM{F}^{\herm}\Ri\bM{V}=\bM{F}^{\herm}\bM{I}_{\T{V}}$ represents the radiation of the currents $\bM{I}_{\T{V}}$ optimal for extinction of power for illumination $\bM{V}$. These are typically not correlated for $\bM{F}\neq \bM{V}$, producing a small contribution from the mixed term and a scattered power close to the lower value in~\eqref{eq:CSg}. The correlated case $\wbM{F}=\wbM{V}$ produce the upper value in~\eqref{eq:CSg}. 

This case shows that an explicitly synthesized beyond diagonal matching or material maximizes the extinction cross section and reaches the limits in~\cite{Gustafsson+etal2020}. Realization of these networks as a material is naturally very challenging, but might at least theoretically be done in a homogenization limit. The results differ from the explicit material synthesis in~\cite{Gustafsson2024b}, which produces optimal local material models for specific excitations.     
 
\section{Far-fields and plane wave excitation}\label{S:PlanewaveScatt}
Interpretations simplify for plane wave excitations and far-field observations. Consider an incident plane wave in the direction $\UV{k}$ and polarization $\UV{e}$. The scattered field is observed in the direction $\UV{r}$ for the polarization $\UV{d}$. 
The scattered power $U$ is normalized to the bi-static radar cross section (RCS)~\cite{Knott+Shaeffer+Tuley2004} 
\begin{equation}
    \sigma_\T{b}(\UV{r},\UV{d},\UV{k},\UV{e}) = \frac{8\pi\eta_0 U}{|E_0|^2}
\end{equation}
with $U$ from~\eqref{eq:maxUnormMoM} and constrained as in~\eqref{eq:CSg}. The quadratic forms are related to the maximum extinction cross section~\eqref{eq:MaxPt}
\begin{equation}
    \max \sigma_\T{t}(\UV{k},\UV{e}) 
    =\wbM{V}^\herm\Ri\wbM{V}
    \to 
    \begin{cases}
            2A(\UV{k}) & \text{sheet} \\
            4A(\UV{k}) & \text{volumetric}         
    \end{cases}
    \label{eq:SigmaExtAsy}
\end{equation}
and similar for $\wbM{F}$. The extinction cross section is further related to the geometrical cross section, or shadow area, $A(\UV{k})$ for electrically large structures. In this regime, the classical limit of $2A$ is observed for thin sheets, where the induced currents radiate bi-directionally, \ie symmetrically away from the sheet. This behavior is associated with shadow scattering or the extinction paradox~\cite{Peierls1979}.
For volumetric objects with low losses, the maximal extinction cross section arises from a forward-scattered field that is out of phase with the incident field, leading to an asymptotic limit $\max\sigma_\T{t} \approx 4A$.
The asymptotic relation $\wbM{V}^{\herm}\Ri\wbM{V} \sim 2A$ also simplifies the application of the Cauchy–Schwarz inequality~\eqref{eq:CSg} 
\begin{equation}
    \frac{\pi}{\lambda^2}A(\UV{k})A(\UV{r})
    \leq \sigma_\T{b}(\UV{r},\UV{d};\UV{k},\UV{e})
    \leq  \frac{4\pi}{\lambda^2}A(\UV{k})A(\UV{r}).
    \label{eq:CSpw}
\end{equation}

Forward scattering $\wbM{F}=\wbM{V}$ yields the forward RCS
$\sigma_\T{b}(\UV{k},\UV{e},\UV{k},\UV{e})=\frac{\pi}{\lambda^2}(\wbM{F}^\herm\Ri\wbM{F})^2=\pi\lambda^{-2}\max\sigma_\T{t}^2$ proportional to the squared maximum extinction cross section~\eqref{eq:MaxPt}. Using the asymptotic expression~\eqref{eq:SigmaExtAsy} for a thin sheet $\sigma_\T{b}\to 4\pi A^2(\UV{k})/\lambda^2$ is recognized as the (monostatic) RCS of a planar object due to the symmetric scattering of the thin sheet. Volumetric objects have 4 times higher forward RCS.     

The non symmetric term $\wbM{F}^\herm\Ri\wbM{V}$ is generally much smaller than the symmetric terms which can be partly understood by identifying $\wbM{F}^\herm\Ri\wbM{V}=\wbM{F}^{\herm}\M{I}_{\T{V}}$ with the radiation towards $\wbM{F}$ of current optimized for radiation towards $\wbM{V}$, \ie the currents in~\eqref{eq:IoptF}. The optimal current for these cases is $\bM{I}_{\T{V}}=\Ri\bM{V}$, which is not correlated with radiation in the $\UV{r}\neq\UV{k} $ direction, but side-lobes can contribute. 

\begin{figure}
    {\centering
    \begin{tikzpicture}
        \def\l{1}
        \def\h{4mm}
        \def\g{1.7}
        \def\d{0.3}
        \def\t{0.1}
        \def\a{120}
        \def\at{300}
        \def\ar{60}
        \def\b{30}
        \begin{scope}
            \draw[thick] (-\l,0) -- (\l,0);
            \draw[<-,dblue] (-\d,\d) -- +(\a:1);
            \draw[->,dred] (\d,\d) -- +(\b:1);
            \draw[->,dashed,dblue] (0,-\d) -- +(\at:1) node[right]{$2A$};            \draw[->,dashed,dblue] (0,\d) -- +(\ar:1) node[right]{$2A$};
            \node at (-1.4,1.4) {(a)};
        \end{scope}
        \begin{scope}[xshift=4cm]
            \draw[thick] (-\l,\t) -- (\l,\t);
            \draw[thick] (-\l,-\t) -- (\l,-\t);
            \draw[<-,dblue] (-\d,\d) -- +(\a:1);
            \draw[->,dred] (\d,\d) -- +(\b:1);
            \draw[->,dashed,dblue] (0,-\d) -- +(\at:1) node[right]{$4A$};
\draw[->,dashed,dblue] (0,\d) -- +(\ar:1) node[right]{$0$};            
            \node at (-1.4,1.4) {(b)};
        \end{scope}
        \begin{scope}[yshift=-3.5cm]
        \begin{scope}[yshift=\h]            
            \draw[thick] (-\l,0) -- (\l,0);
            \draw[<-,dblue] (-\d,\d) -- +(\a:1);
            \draw[->,dred] (\d,\d) -- +(\b:1);
\draw[->,dashed,dblue] (0,\d) -- +(\ar:1) node[right]{$2A$};            
         \end{scope}
         \draw[very thick,gray] (-\g,0) -- (\g,0);
        \begin{scope}
            \draw[->,densely dashed,dblue] (\d,-\d) -- +(\at:1) node[right]{$2A$};
         \end{scope}
        \node at (-1.4,1.4) {(c)};
        \end{scope}
        \begin{scope}[xshift=4cm,yshift=-3.5cm]
        \begin{scope}[yshift=\h]            
            \draw[thick] (-\l,\t) -- (\l,\t);
            \draw[thick] (-\l,-\t) -- (\l,-\t);
            \draw[<-,dblue] (-\d,\d) -- +(\a:1);
            \draw[->,dred] (\d,\d) -- +(\b:1);
\draw[->,dashed,dblue] (0,\d) -- +(\ar:1) node[right]{$2A$};            
         \end{scope}
         \draw[very thick,gray] (-\g,0) -- (\g,0);
        \begin{scope}
            \draw[->,densely dashed,dblue] (\d,-\d) -- +(\at:1) node[right]{$2A$};
         \end{scope}
        \node at (-1.4,1.4) {(d)};
        \end{scope}
    \end{tikzpicture}\par}
    \caption{Examples of scattering configurations (side view) analyzed in this paper. (a) single meta surface. (b) multiple layers. (cd) ground plane below the ab) structures. The asymptotic forward scattering is indicated, with $A$ denoting the shadow area. In this paper, rectangular regions with side lengths $\ellx=2\elly$ are used as illustrated in Fig.~\ref{fig:antennascattreg}.}
    \label{fig:ScattConfig}
\end{figure}
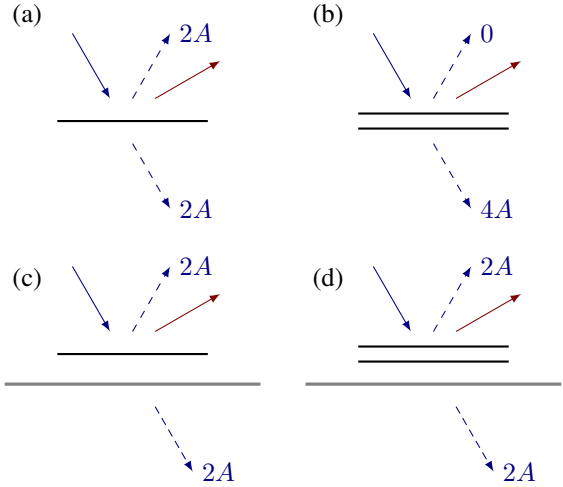

Many relevant scattering geometries can be analyzed using the proposed technique. Here, we focus on the four configurations illustrated in Fig.~\ref{fig:ScattConfig}. The first case, a), consists of a thin sheet modeled by a surface resistivity~\cite{VanBladel2007}. The thin sheet scatters symmetrically (bidirectional) in the upward and downward directions. Asymptotically, this results in a forward scattering cross section of approximately $2A$, \cf~\eqref{eq:CSpw}, with a similar contribution in the specular direction. The cross term is small outside the forward and specular directions. Consequently, this leads to an approximation based on the upper limit in~\eqref{eq:CSpw} in these directions (and co-polarization), and to the lower limit outside them. The resulting factor of 4 (or $-6\unit{dB}$) is consistent with~\cite{Abdelrahman+Monticone2023}.

Using a volumetric region or multiple layers, as shown in Fig.~\ref{fig:ScattConfig}b, breaks the up- and down-symmetry. At the same time, the maximal forward scattering doubles to $4A$, \cf~\eqref{eq:CSpw}. However, the cross term generally remains small except in the forward direction, leading to an approximate realization of the lower limit in~\eqref{eq:CSpw}, but with $A$ replaced by $2A$. Hence, this produces a bistatic RCS limit similar to the single-layer case, but valid in all directions.

In the third and fourth cases, Fig.~\ref{fig:ScattConfig}c and Fig.~\ref{fig:ScattConfig}d, a ground plane is added to the previous configurations. These cases can be analyzed by introducing a mirror incident field~\cite{Gustafsson+Sjoberg2011}. This symmetric illumination results in an asymptotic forward scattering of $\wbM{V}^{\herm}\Ri\wbM{V} \approx 2A$ for both cases, corresponding to a strong specular reflection. The mixed term remains small, again leading to an asymptotic result similar to case a).

Note that these conclusions are based on asymptotic results for electrically large structures and serve as a complement to numerical simulations. The observation that the single-layer and ground-plane configurations are limited to approximately a factor of 4 (or $-6\unit{dB}$) in scattering intensity for anomalous (non-specular) directions compared to the forward or specular direction is interesting, and propose a simple estimate of the realizable scattering in anomalous directions as a quarter of the scattering in the forward (or specular) direction. Earlier investigations indicate that achieving perfect anomalous reflection is challenging~\cite{Asadchy+etal2017,Mohammadi+Alu2016} and a $-6\unit{dB}$ is derived for thin sheets in~\cite{Abdelrahman+Monticone2023}.

Case (b), involving multiple layers or a volumetric structure, is more surprising, as the strong anomalous scattering is accompanied by strong forward scattering. The asymptotic forward scattering of $4A$, twice the shadow scattering~\cite{Peierls1979} (also known as the extinction paradox), corresponds to a scattered field that is out of phase with the incident field. Simultaneously, this forward-scattered field is accompanied by scattering of similar intensity as the specular reflection from a plate, but directed anomalously. It is important to note here that power is conserved (the foundation of the derivation, see discussion in~\cite{Gustafsson+etal2020}), and that such scattering behavior can be realized using an idealized passive, yet non-local, material model (i.e., a matching network).

\section{Numerical examples}\label{S:NumRes}
\begin{figure}
    \centering
    \includegraphics[width=0.95\linewidth]{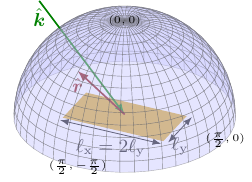}
    \caption{Illustration of the scattering setup with an illuminating plane wave from in the $\UV{k}$ direction and maximization of the scattered field in the $\UV{r}$ direction. The directions $\UV{k}$ and $\UV{r}$ are represented in a spherical coordinate system $(\theta,\phi)$ as indicated in the figure.
    In the presented numerical examples, a rectangular resistive sheet with surface resistance $R_\T{s}=0.01\unit{\Omega/\square}$ and side lengths $\ellx=2\elly$ is used in the configurations depicted in Fig.~\ref{fig:ScattConfig}. }
    \label{fig:scattgeo}
\end{figure}

The maximal scattered power in the far-field from an incident plane wave is described by the bi-static radar cross section. Consider planar rectangular regions, as depicted in Fig.~\ref{fig:ScattConfig} and Fig.~\ref{fig:scattgeo}. The rectangle is the antenna design region and models the antenna structure, \eg a patch array as in Fig.~\ref{fig:antennascatt}, by only keeping its material losses, here modeled by a surface resistivity $\Rs$, \eg $\Rs\approx 0.01\unit{\Omega/\square}$ for Cu around $1\unit{GHz}$ used though out the examples. The sheets are separated $\lambda/4$ for in (bd) cases and placed $\lambda/4$ above the ground plane for cases (cd). The problem is solved by evaluating the matrices $\M{R},\M{V},\M{F}$ in~\eqref{eq:maxUMoM}, see the appendix. Numerical investigations reveal that it is sufficient to use around 8 discretization elements per wavelength for the considered cases. Note that the specular reflection from the ground plane is not included in the plots. 

\begin{figure}
    \centering
    \begin{tikzpicture}
    \node at (0,0) {\includegraphics[width=0.45\linewidth]{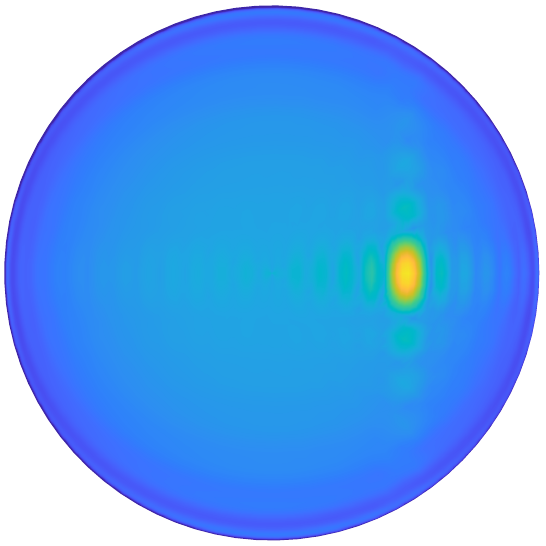}};
    \node at (-1.8,2) {(a)};
    \node at (4.5,0) {\includegraphics[width=0.45\linewidth]{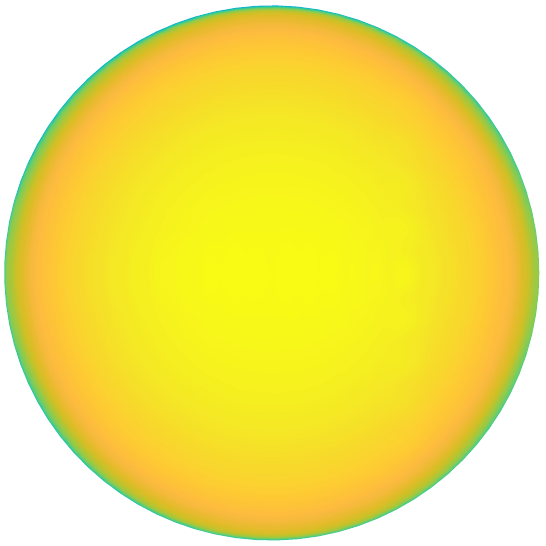}};
    \node at (3.2,2) {(b)};
    \node at (0,-4.5) {\includegraphics[width=0.45\linewidth]{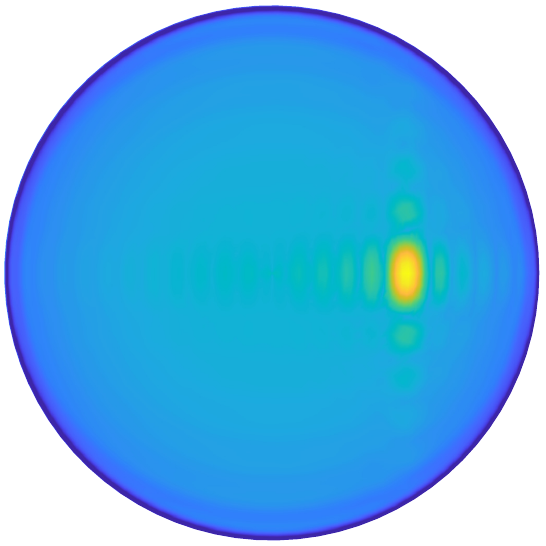}};
    \node at (-1.8,-2.5) {(c)};
    \node at (4.5,-4.5) {\includegraphics[width=0.45\linewidth]{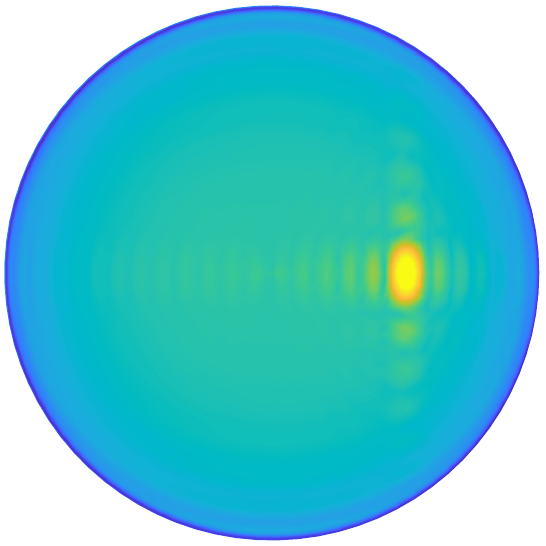}};
    \node at (3.2,-2.5) {(d)};  \node at (2.1,-2.5) {\includegraphics[width=0.13\linewidth]{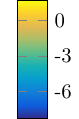}};  
\end{tikzpicture}
    \caption{Upper bound on the bi-static scattering $\sigma_\T{b}$ in dB for the four configurations in Fig.~\ref{fig:ScattConfig} for a 2:1 rectangular region with longest side length $\ellx=10\lambda$, see Fig.~\ref{fig:scattgeo}, modeled with surface resistivity $R_\T{s}=0.01\unit{\Omega/\square}$. Plotted by projecting the upper hemisphere in Fig.~\ref{fig:scattgeo} on the xy-plane.} 
    \label{fig:scattplot3D_ka35}
\end{figure}
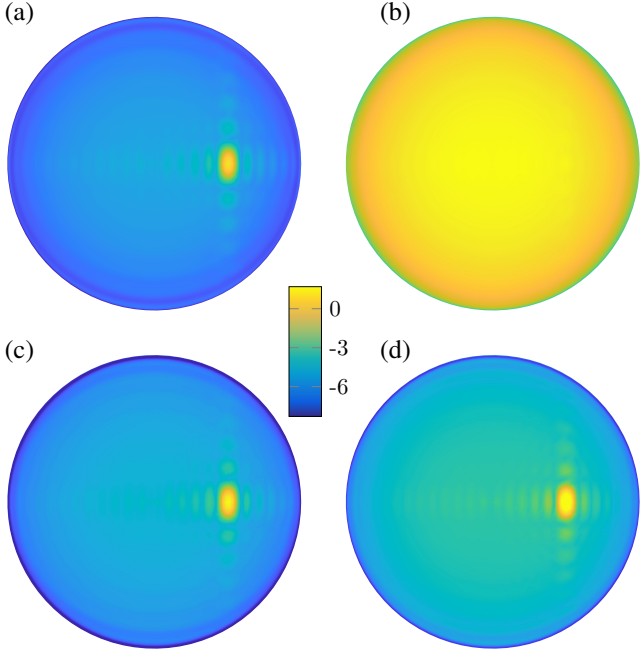

The four scattering configurations in Fig.~\ref{fig:ScattConfig}, applied to the $10 \times 5\lambda^2$ rectangular sheets shown in Fig.~\ref{fig:scattgeo}, are illustrated in Fig.~\ref{fig:scattplot3D_ka35}. An incident plane wave is assumed, arriving from $\theta = 30^\circ$, $\phi = 180^\circ$ (equivalently, $\theta = -30^\circ$, $\phi = 0^\circ$), with TM polarization (i.e., $\UV{\theta}$-polarized). The results in Fig.~\ref{fig:scattplot3D_ka35} show the maximal bistatic RCS, $\sigma_\T{b}$, over the upper hemisphere ($|\theta| \leq 90^\circ$), projected onto the $xy$-plane. The RCS is normalized by $4\pi A^2/\lambda^2$ with $A=\ellx\elly$.

The single-sheet configuration in Fig.~\ref{fig:scattplot3D_ka35}a shows a high RCS in the specular direction, with significantly lower values (approximately $-6\unit{dB}$) in other directions. Additionally, the RCS decreases near grazing angles, i.e., for $\theta \approx 90^\circ$.

The bi-directional scattering symmetry is broken in the two-sheet (or multi-layer) configuration of Fig.~\ref{fig:ScattConfig}b, as illustrated in Fig.~\ref{fig:scattplot3D_ka35}b. In this case, the RCS appears nearly uniform across all directions, with no pronounced enhancement in the specular direction. However, the RCS amplitude remains comparable to the specular peak of the single-sheet case, which is consistent with the discussion in Sec.~\ref{S:PlanewaveScatt}.

The ground-plane configurations shown in Fig.~\ref{fig:scattplot3D_ka35}c and Fig.~\ref{fig:scattplot3D_ka35}d exhibit scattering behavior similar to that of the single-sheet case in Fig.~\ref{fig:scattplot3D_ka35}a. This similarity is explained by the bi-directional scattering characteristics of the configurations in Fig.~\ref{fig:ScattConfig}.

\begin{figure}
    \centering
    \includegraphics[width=0.95\linewidth]{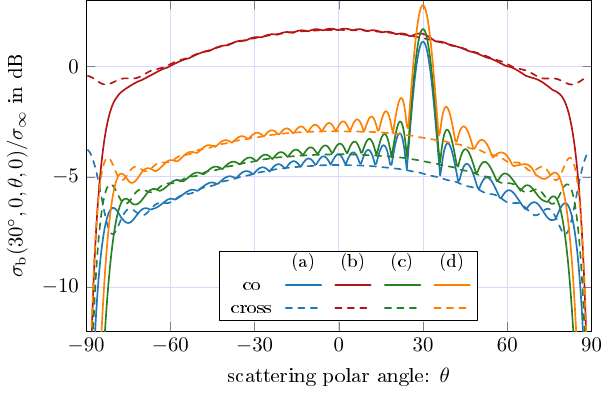}
    \caption{Normalized maximal RCS $\sigma_\T{b}$ for the scattering configurations in Fig.~\ref{fig:ScattConfig} and Fig.~\ref{fig:scattgeo} with surface resistivity $R_\T{s}=0.01\unit{\Omega/\square}$ and incident TM polarized plane wave from $\theta=-30^{\circ}$, and electrical size $\ellx=10\lambda$. The RCS is normalized by the monostatic RCS $\sigma_\infty=4\pi A^2/\lambda^2$. }
    \label{fig:scattplot2D_ka35}
\end{figure}

The 2D plots in Fig.~\ref{fig:scattplot3D_ka35} are complemented by restricting the scattering analysis to the $xz$-plane (i.e., $\phi = 0^\circ$ or $\phi = 180^\circ$), as shown in Fig.~\ref{fig:scattplot2D_ka35}. Cross-polarized results are also included in the figure. As before, cases (a), (c), and (d) exhibit similar scattering patterns, with only slight amplitude variations, case (d) showing the highest RCS in the specular direction. In contrast, case (b) does not display a distinct specular reflection, instead exhibiting a relatively uniform RCS across angles. The co- and cross-polarized results are also largely similar, except in the specular directions and near grazing angles.

\begin{figure}
    \centering
    \includegraphics[width=0.95\linewidth]{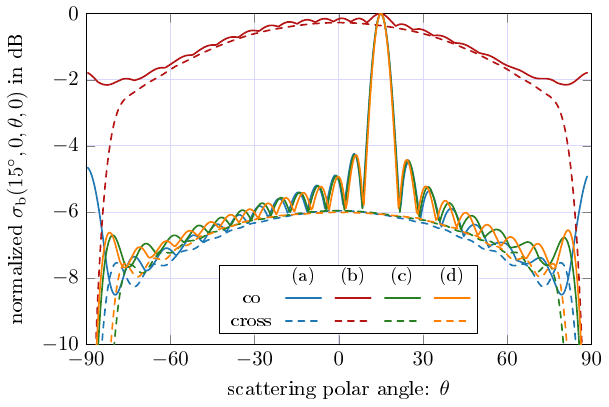}
    \caption{Normalized maximal RCS $\sigma_\T{b}$ for the scattering configurations in Fig.~\ref{fig:ScattConfig} and Fig.~\ref{fig:scattgeo} with surface resistivity $R_\T{s}=0.01\unit{\Omega/\square}$, incident TE polarized plane wave from $\theta=-15^{\circ}$, and electrical size $\ellx=8\lambda$. The RCS is normalized by its maximum value.}
    \label{fig:scattplot2D_ka52}
\end{figure}

The results are also similar for a TE-polarized (i.e., $\UV{\phi}$) incident plane wave, as shown in Fig.~\ref{fig:scattplot2D_ka52}. In this case, the incident wave arrives from $\theta = -15^\circ$, and the structure has an electrical length of $\ellx = 8\lambda$. To facilitate comparison, the results are normalized by their respective maximum values, which leads to nearly overlapping curves for cases (a), (c), and (d).

We note that the quadratic forms $\wbM{F}^\herm\Ri\wbM{F}$ (or equivalently, $\wbM{V}^\herm\Ri\wbM{V}$) and $\wbM{F}^\herm\Ri\wbM{V}$ are the key quantities of interest for evaluating the maximal bistatic RCS~\eqref{eq:maxUnormMoM}. The symmetric term $\wbM{F}^\herm\Ri\wbM{F}$ corresponds to the maximal extinction cross section~\cite{Gustafsson+etal2020}, with an asymptotic limit proportional to the geometrical cross-sectional area (or, more generally, the shadow area) of the region~\eqref{eq:SigmaExtAsy}.

\begin{figure}
    \centering
    \includegraphics[width=0.49\linewidth]{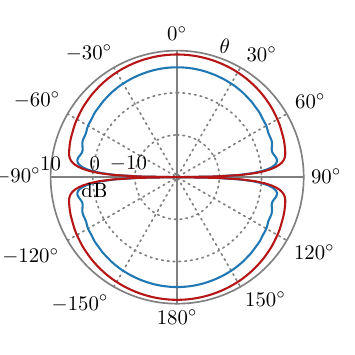}
    \includegraphics[width=0.49\linewidth]{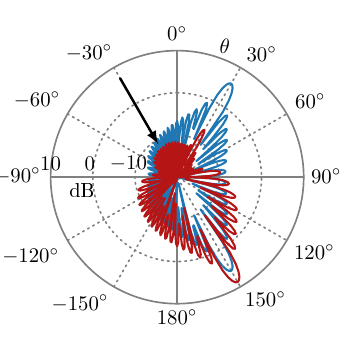}
    \caption{Illustration of the terms (left) $\wbM{F}^\herm\Ri\wbM{F}$ and (right) $\wbM{F}^\herm\Ri\wbM{V}$ for the (a) in blue and (b) in red cases in Fig.~\ref{fig:ScattConfig} and Fig.~\ref{fig:scattgeo} with $\ellx=10\lambda$ and $R_\T{s}=0.01\unit{\Omega/\square}$. The excitation to the right is from a TM plane wave from $\theta=-30^{\circ}$ and $\phi=0$ as illustrated by the arrow and all quantities are plotted for the (co-polarized) $\UV{\theta}$ component at $\phi=0$.} 
    \label{fig:scattplotPolar_ka35}
\end{figure}

The term $\wbM{F}^\herm\Ri\wbM{F}$ is shown for the two configurations in Fig.~\ref{fig:ScattConfig}a and Fig.~\ref{fig:ScattConfig}b in Fig.~\ref{fig:scattplot2D_ka35}, for an electrical size of $\ellx = 10\lambda$. It is observed that the extinction cross section for case (b) is approximately twice as large (i.e., $3\unit{dB}$ higher) as that of case (a). This can be explained by the symmetric scattering of the thin sheet in case (a), which radiates equally in the upward and downward directions, thereby reducing the forward scattering, as discussed in Sec.~\ref{S:PlanewaveScatt}. The variation of scattering with respect to the illumination angle is relatively weak and is mainly attributed to the reduced cross-sectional (shadow) area at oblique angles. Additionally, the bistatic scattering vanishes at $\theta = \pm 90^\circ$ for the depicted TM polarized case.

Using the optimal currents $\bM{I}_\T{o} = \Ri\wbM{F}$~\eqref{eq:IoptF} allows us to reinterpret the forward scattering term $\wbM{F}^\herm\Ri\wbM{F}$ as the radiation $\wbM{F}^\herm\bM{I}_\T{o}$ generated by the optimal current. This radiation is strongest in the forward direction, which the optimal current is designed to maximize. The mixed term $\wbM{F}^\herm\Ri\wbM{V}$ represents the radiation in directions other than the forward one. In most of these directions, the optimal currents do not interfere constructively, resulting in weaker radiation.

The radiation pattern corresponding to the mixed term $\wbM{F}^\herm\Ri\wbM{V}$ is shown in the right panel of Fig.~\ref{fig:scattplotPolar_ka35} for the configurations in Fig.~\ref{fig:ScattConfig}ab. For the volumetric object (b), a strong main lobe is observed in the forward direction, with weaker side lobes elsewhere. This suggests that the mixed term contributes significantly only in the forward direction and can be neglected elsewhere. In contrast, the thin sheet in case (a) exhibits two main lobes due to its symmetric radiation pattern. As a result, the mixed term also contributes in the specular direction for this configuration.
Although the sidelobes are weak for the considered cases, periodic regions have grating lobes with equal strength to the main lobe and contribute. 

\begin{figure}
    \centering
    \includegraphics[width=0.95\linewidth]{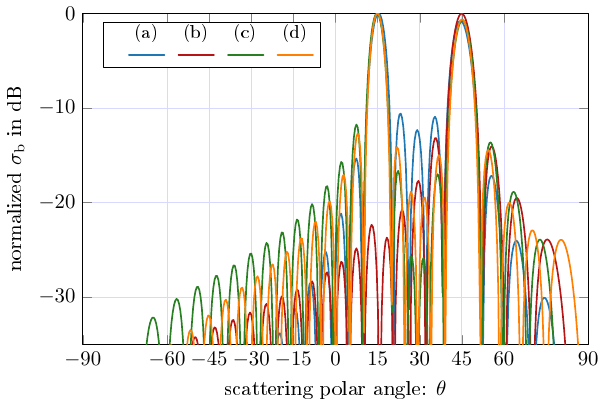}
    \caption{Normalized RCS $\sigma_\T{b}$ for the scattering configurations in Fig.~\ref{fig:ScattConfig} and Fig.~\ref{fig:scattgeo} with surface resistivity $R_\T{s}=0.01\unit{\Omega/\square}$ and incident TM polarized plane wave from $\theta=-15^{\circ}$ maximized for radiation in the $\theta=45^\circ$-direction, and electrical size $\ellx=10\lambda$. The RCS is normalized by its maximal value. }
    \label{fig:scattplot2D_ka35_theta}
\end{figure}

Normalized RCS for a scatterer synthesized using the procedure outlined in Sec.~\ref{S:PlanewaveScatt}, with an illuminating wave incident from $\theta=15^\circ$ and $\phi=180^\circ$ and optimized for scattering toward $\theta=45^\circ$ and $\phi=0^\circ$, is depicted in Fig.~\ref{fig:scattplot2D_ka35_theta}. The RCS is shown for the four configurations in Fig.~\ref{fig:ScattConfig}, using a rectangular resistive design region from Fig.~\ref{fig:scattgeo} with $\ellx=10\lambda$.

The three cases (a, c, and d) exhibit a specular reflection at $\theta=15^\circ$ with an amplitude comparable to the desired anomalous scattering at $\theta=45^\circ$. In contrast, case (b) shows negligible specular scattering and instead produces a focused beam directed at $\theta=45^\circ$.

It is important to emphasize that, although the fundamental bounds are unique, the computed solutions may not be. Multiple current distributions can achieve similar scattering performance in the preferred direction while exhibiting different behavior in other directions. If desired, additional constraints can be incorporated into the optimization problem to suppress specular scattering or tailor the scattering characteristics more precisely.

\section{Conclusions}\label{S:Conclusions}
Fundamental limits on the anomalous scattering of metasurfaces have been presented. The derived bounds apply to all metasurfaces constructed from materials with resistivity above the threshold value used in the bound calculations. General electrical materials, such as metals and dielectrics, are included in the model, whereas magnetic materials, modeled through general bi-anisotropic permittivity and permeability tensors~\cite{VanBladel2007}, are not considered. However, metamaterials homogenized from microscopic structures composed of conductors and dielectrics are included in the analysis. Extending the framework to incorporate magnetic materials is an important direction for future work, as many metasurface synthesis techniques rely on such properties.

An explicit synthesis method for non-local (beyond-diagonal) matching networks has also been presented, generalizing the explicit local material synthesis approach developed in~\cite{Gustafsson2024b} to the broader case of bounds based on power conservation principles.

The results are valuable for the understanding and design of reflector arrays, metasurface antennas, and reconfigurable intelligent surfaces (RIS). Arbitrary configurations are easily addressed numerically, while analytical asymptotic results for bistatic radar cross sections (RCS) complement the numerical studies and provide physical insight. In particular, the asymptotic analysis predicts a typical $6\unit{dB}$ reduction in anomalous scattering relative to forward (or specular) scattering for single sheets and structures placed above a ground plane.

The results also apply to the maximization of power in the near field or in a specific radiated field. Generalizations to multiple beams are of significant interest for RIS applications, where multiple degrees of freedom are desired. Furthermore, additional constraints such as sidelobe suppression and control of specular scattering can be incorporated into future investigations.

\appendices
\section{MoM modeling}\label{S:MoM}
Throughout this paper it is assumed that a sufficiently fine mesh is used such that the numerical discretizations errors can be neglected.
The excitation matrix $\M{V}$ is defined by the projection of the incident field $\V{E}_\T{in}(\V{r})$ on the used basis functions $\basv_n$
\begin{equation}
    V_n = \int\basv_n(\V{r})\cdot\V{E}(\V{r})\diffV.
\end{equation}
The coefficients $V_n$ are collected in the column matrix $\M{V}$. 
For an incident plane wave $\V{E}(\V{r})=E_0\UV{e}\eu^{-\ju k\V{r}\cdot\UV{k}}$ we have
\begin{equation}
    V_n = E_0\int\basv_n(\V{r})\cdot\UV{e}\eu^{-\ju k\V{r}\cdot\UV{k}}\diffV.
\end{equation}

The corresponding radiated field defines the matrix $\M{F}$ which in the far-field for a direction $\UV{r}$ and polarization $\UV{e}$ simplifies to
\begin{equation}
    F_n = \ju\frac{\sqrt{\eta_0} k}{4\pi}\int\basv_n(\V{r}_1)\cdot\UV{e}\eu^{-\ju k\UV{r}\cdot\V{r}_1}\diffV_1.
\end{equation}
The MoM impedance matrix $\M{Z}$ is defined as in standard EFIE-based MoM formulations~\cite{Harrington1968}, see also~\cite{Gustafsson+etal2016a,Gustafsson+etal2020}. 

\section{QCQP solution}\label{S:QCQP}
The solution of the dual problem~\eqref{eq:dualB} is obtained for the Lagrange parameter
\begin{equation}
\nu_\T{o}=\M{F}^{\herm}\Ri\M{F}+|\M{F}^{\herm}\Ri\M{V}|\sqrt{\frac{\M{F}^{\herm}\Ri\M{F}}{\M{V}^{\herm}\Ri\M{V}}}.
\label{eq:dualpar}
\end{equation}
The corresponding optimal current is
\begin{equation}
    \M{I}_\T{o} 
    = \frac{\nu_\T{o}}{2}(\nu_\T{o}\M{R}-\M{F}\M{F}^{\herm})^{-1}\M{V}
    =\frac{1}{2}\Ri\M{V} + \frac{\alpha}{2}\Ri\M{F}.
\end{equation}
using the Sherman-Morrison formula~\cite{Golub+Loan2013}. The weight is 
\begin{equation}
   \alpha =\frac{\M{F}^{\herm}\Ri\M{V}}{|\M{F}^{\herm}\Ri\M{V}|}\sqrt{\frac{\M{V}^{\herm}\Ri\M{V}}{\M{F}^{\herm}\Ri\M{F}}}
\end{equation}



\end{document}